\documentclass{article}
\usepackage{spconf,amsmath,graphicx}
\usepackage{hyperref}
\usepackage{multirow}
\usepackage{adjustbox}
\usepackage{graphicx}
\title{multimodal transformers for utterance-level code-switching detection}
\name{Krishna D N}
\address{Freshworks Inc.}
\begin{document}
%\ninept
%
\maketitle
\begin{abstract}
An utterance that contains speech from multiple languages is known as a code-switched sentence. In this work, we propose a novel technique to predict whether given audio is monolingual or code-switched. We propose a multi-modal learning approach by utilizing the phoneme information along with audio features for code-switch detection.
Our model consists of a Phoneme Network (PN) that processes phoneme sequence and Audio Network (AN), which processes the MFCC features. We fuse representation learned from both the Networks to predict if the utterance is code-switched or not. The Audio Network and Phonetic Network consist of initial convolutional, Bi-LSTM, and transformer encoder layers. The transformer encoder layer helps in selecting important and relevant features for better classification by using self-attention. We show that utilizing the phoneme sequence of the utterance along with the MFCC features improves the performance of code-switch detection significantly. We train and evaluate our model on Microsoft's code-switching challenge \href{https://www.microsoft.com/en-us/research/event/workshop-on-speech-technologies-for-code-switching-2020/}{dataset} for Telugu, Tamil, and Gujarati languages. Our experiments show that the multi-modal learning approach significantly improved accuracy over the uni-modal approaches for Telugu-English, Gujarati-English, and Tamil-English datasets. We also study the system performance using different neural layers and show that the transformers help obtain better performance.
\end{abstract}
\begin{keywords}
multimodal learning, transformers, code-switching
\end{keywords}
\section{Introduction}
\label{sec:intro}
A continuous change between two or more languages within a single utterance is known as code-switching [1]. Code-switching detection is about identifying whether an utterance contains multiple languages. Low resource languages like Indian languages most commonly have code-switches with  English. Code-switched utterances usually cause problems for automatic speech recognition systems, which are built using monolingual data only. Many researchers in the speech community are trying to develop systems that can handle code-switching speech. Robust acoustic-modeling using Bi-lingual Deep Neural Networks are used in [2] for Frisian code-switched automatic speech recognition task. [3] proposes to use a shared acoustic model between the code-switched languages to obtain a robust German speech recognition system. [4] proposes a one-pass recognition strategy and shows we can eliminate the classical multi-pass technique, which follows language identification, language boundary detection, and language-independent speech recognition. Recently,[5] proposes using an end-to-end neural network framework that uses the attention-CTC-based seq2seq model. They propose to jointly predict the language label and the CTC labels for robust code-switched speech recognition for Chinese. Language modeling techniques have also been used for code-switched speech recognition tasks recently [7]. [6] proposed to use a recurrent neural network to predict the code-switches by using textual features like Part of speech. Language identification seems to be one of the main components of code-switch identification and code-switched speech recognition. It is shown to have provided good improvement in the system performance [8,9,10,11]. Some interesting works[1,2,3,4] have shown different methods to detect the code-switching in a single conversation/utterance. Multilingual training of DNN-based speech recognition systems has shown some improvements in the low resource languages[13,14,15,16]. Various other techniques like multilingual DNNs[12], using latent space language models[13], using untranscribed data[14], and data augmented acoustic and language models [15] are also proposed to code-switch detection.

Recent developments in the area of deep learning have shown tremendous improvement in many areas of speech processing, including speech recognition [16,17,18,19,20], language identification [21,22,23,24], speaker identification [25,26,27], emotion recognition [28,29,30], and many others. Recently, the Attention mechanism [31] is shown to be a dominating approach in both Natural language processing and speech due to their ability to focus on specific parts of the input signal to extract relevant information for a particular task. Sequence to sequence models become more powerful and efficient when it is built using just attention layers. Work by [32] builds an end-to-end architecture using many attention layers, thus eliminating the need for a convolutional network or LSTM layers. These models are popularly known as Transformer, and they are the state of the art models for most Natural language processing problems. In the speech recognition field, state of the art speech recognition uses attention based sequence to sequence model due to its efficiency and capacity to align and predict. These models are also used extensively in problems like emotion recognition [28], language identification, and so on.

It is shown that the language identification problem can be tackled by using phoneme sequences extracted from the speech recognition system with some post processing[ 38]. [38],describes that we can use an English phoneme recognition system to extract phoneme sequences for all the utterances with a different language and apply some N-gram statistics post-processing to classify test sentences. Motivated by the work of M. A. Zissman et al. [38], we propose to use phoneme sequences as an additional guiding signal along with MFCC features for the network architecture. We propose a multi-modal transformer model for utterance-level code-switch detection. Our model consists of 2 streams, Audio Network (AN) and Phoneme Network (PN). The Audio Network consists of a sequence of CNN and LSTM layers followed by a Transformer encoder to process MFCC features. The Phoneme Network also consists of a sequence of CNN and LSTM layers followed by a Transformer Encoder layer to process phoneme embeddings. We pool the output features from transformer layers from both Audio Network and Phoneme Network using the statistics pooling layer, and we concatenate these features to create utterance-level multi-modal feature representation. We project this multi-modal feature representation to lower dimensional space using a fully connected layer. Finally, we classify the lower-dimensional representation using a softmax layer to predict the class label. We conduct all our experiments on the Microsoft code-switching challenge dataset for three Indic languages Telugu-English, Tamil-English, and Gujarati-English. Our experimental study shows that our approach significantly improves the system performance compared to the baseline model. We also show that our multi-modal approach outperforms the uni-modal methods.
Our open source implementation can be found in \url{https:/github.com/KrishnaDN/Code-Switch-Detection}

The organization of the paper is as follows. In section 2, we explain our proposed approach in detail. In section 3, we give a detailed analysis of the dataset, and in section 4, we explain our experimental setup in detail. Finally, in section 5, we describe our results.

\section{Multi-Modal Learning}
\label{sec:majhead}
This paper proposes a multi-modal learning approach for utterance-level code-switch detection using audio features and phoneme information. The model consists of 2 stream architecture containing Audio Network and Phoneme Network, as shown in figure 1. 
The Audio network takes MFCC features as input, as shown in Fig 1. The Audio Network consists of Audio Encoder, Audio Transformer, and statistics pooling blocks. The Audio Encoder converts the MFCC features into high-level representations using Convolutional neural networks followed by a Bi-LSTM layer, as shown in Fig. 1. This high-level feature representation is fed into the Audio Transformer block, consisting of 2 multi-head self-attention layers. The multi-head self-attention layers use self-attention to select important and relevant features for code-switching detection using attention weighting. 
The statistics pooling layer in Audio Network helps in aggregating frame-level features from Audio Transformer into an utterance-level feature vector. Similarly, the Phoneme Network consists of Phoneme Encoder, Phoneme Transformer, and statistic pooling layer, as shown in Fig. 1. The Phoneme Encoder takes phoneme embeddings as input and produces high-level contextual representation using convolutional neural networks and Bi-LSTM. These features are processed by Phoneme Transformer to select relevant features for better code-switch classification using multi-head self-attention layers. The Phoneme network's statistics-pooling layer acts as an aggregation layer to pool variable length feature vectors from Phoneme Transformer into a single feature vector for classification.
We use Early fusion to concatenate the utterance-level feature vector from both the Audio Network and Phoneme Network to combine information from both audio features and phoneme sequences for better classification performance. We explain all the blocks in detail in the following section.

\begin{figure}[t]
  \centering
  \includegraphics[width=\linewidth]{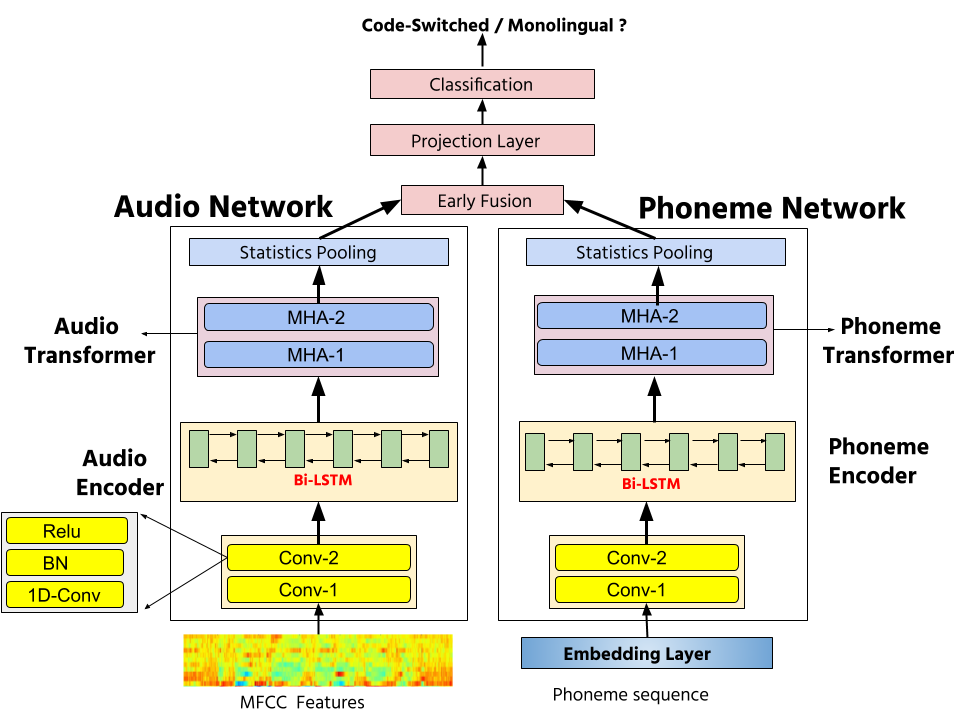}
  \caption{Proposed model architecture.}
  \label{fig:speech_production}
\end{figure}

\subsection{Phoneme Recognition}
Our proposed method consists of 2 streams known as Audio Network and Phoneme network. The Phoneme Networks take a sequence of phonemes corresponding to the audio data as its input. Due to the lack of ground truth phoneme labels in the code-switch detection dataset, we generate phoneme labels by training a GMM-HMM based speech recognition model for Telugu, Tamil, and Gujarati Languages using Microsoft's speech recognition challenge \href{https://www.microsoft.com/en-us/research/publication/interspeech-2018-low-resource-automatic-speech-recognition-challenge-for-indian-languages/}{dataset}. Using these speech recognizers for the three languages, we create phoneme labels for Telugu-English, Tamil-English, and Gujarati-English code-switching dataset. These phoneme sequences and their corresponding MFCC features are used during the training of code-switch identification models. The Audio Network takes MFCC features for each audio file, while Phoneme Network takes the phoneme sequence generated by the GMM-HMM speech recognition model for that input audio. All the phoneme recognizers are trained in Kaldi.

\subsection{Audio Encoder}
\label{ssec:subhead}
The Audio Network consists of Audio Encoder, Audio Transformer, and statistics pooling layer, as shown in the figure. The Audio Encoder takes a sequence of MFCC frames as input and produces an utterance-level feature vector. We extract 13-dimensional MFCC feature vectors for every 10ms using a 25ms window. These low-level features are passed as input to Audio Encoder, followed by Audio transformer to learn feature representation, which contains information about code-switch in an utterance.

The Audio Encoder processes audio data to extract features that are useful for code-switch identification. It consists of two 1D convolutional layers followed by a Bi-LSTM layer. The Audio Encoder takes a sequence of 13-dimensional MFCC features and applies 1D convolution operations to extract higher-level feature representations. Each convolutional block consists of a 1D convolution operation followed by Batch normalization and Relu activation, as described in  Fig.1. Each convolutional layer has its kernel size and filters. We have [1x7,1x5] filter size for Conv-1 and Conv-2 layer, respectively. Both Conv-1 and Conv-2 blocks are having stride [1x3]. Similarly, we have 64 filters for both Conv-1 and Conv-2. Since the convolutional layer does not capture temporal information, we use Bi-LSTM right after the convolutional block to capture temporal information.Let $\boldsymbol{X=[x_1,x_2..x_n,...x_T]}$ be a sequence of $\boldsymbol{T}$ MFCC feature vectors of dimension 13 and input vector $\boldsymbol{x_n}$ is a MFCC frame extracted from the raw audio for every 25ms window with 10ms shift.

\begin{equation}
\boldsymbol{F^A} = \textsf{Convolution}(\boldsymbol{X})
\end{equation}

Where \textsf{Convolution} is a sequence of two 1D convolutional blocks applied to the MFCC sequence $\boldsymbol{X}$, as shown in Fig.1. After \textsf{Convolution} operation, we obtain a feature sequence $\boldsymbol{F^A=[f_1,f_2.....f_T\prime_{}]}$ of length $\boldsymbol{T^\prime_{}}$. The feature matrix, $\boldsymbol{F^A}$ has an x-axis as the time dimension, and the y-axis is a feature dimension. The feature dimension size will be the same as the number of filters in the last convolutional layer. In our case, the feature dimension is 64, as the number of filters in the last convolutional layer is 64. Since convolution cannot capture temporal information, we use Bi-LSTM Layers to learn the temporal representation of the feature matrix $\boldsymbol{F^A}$

\begin{equation}
\boldsymbol{H^A} = \textsf{Bi-LSTM}(\boldsymbol{F^A})
\end{equation}

Where \textsf{Bi-LSTM} represents  Bidirectional LSTM layer whose hidden dimension is 128. $\boldsymbol{H^A=[h_1,h_2.....h_T\prime_{}]}$ represents the output sequence from the final Bi-LSTM layer. $\boldsymbol{h_i}$ represents the $i^{\text{th}}$ timestep hidden activity from the last layer. Since the LSTM is operating in both directions, the final hidden layer output feature will be 2 times the LSTM hidden feature vector in size.

\subsection{Phoneme Encoder}
\label{ssec:subhead}
The Phoneme Network consists of Phoneme Encoder, Phone Transformer, and statistics pooling layer. The Phoneme Encoder consists of a learnable phoneme embedding layer, two layers of convolution followed by a Bi-LSTM layer, as shown in Figure 1. The Phoneme Encoder takes a phoneme sequence as inputs to the embedding layers, and these fixed dimensional embeddings are learned during training. The phoneme embeddings from the embedding layer are passed into the convolutional network, consisting of 1D convolution layers. Each convolutional layer has its kernel size and filters. we have [1x3,1x5] filter size for Conv-1 and Conv-2 layer respectively. Similarly, we have 64 filters for both Conv-1 and Conv-2. Since the convolutional layer does not capture temporal information, we use Bi-LSTM right after the convolution block. The Bi-LSTM learns contextual representations in the phoneme sequence and helps in detecting code-switching.Let $\boldsymbol{P=[p_1,p_2..p_n,...p_N]}$ be a sequence of $\boldsymbol{N}$ phoneme embeddings of dimension 128 and input vector $\boldsymbol{p_n}$ is a phoneme embedding for $n^{\text{th}}$ phoneme.

\begin{equation}
\boldsymbol{F^P} = \textsf{Convolution}(\boldsymbol{P})
\end{equation}

Where, \textsf{Convolution} is a sequence of two convolutional blocks applied to the sequence of phoneme embeddings sequence $\boldsymbol{P}$ as shown in Figure 1.  $\boldsymbol{F^P=[f_1,f_2.....f_N\prime_{}]}$ is high-level feature representation learned from convolution operation. Since convolution cannot capture temporal information, we use a sequence of Bi-LSTM Layers to learn a temporal representation of the feature matrix $\boldsymbol{F^A}$. 

\begin{equation}
\boldsymbol{H^P} = \textsf{Bi-LSTM}(\boldsymbol{F^P})
\end{equation}

Where \textsf{Bi-LSTM} represents 2 layers Bidirectional LSTM network whose hidden dimension is 128. $\boldsymbol{H^P=[h_1,h_2.....h_N\prime_{}]}$ represents the output sequence from the final Bi-LSTM layer. $\boldsymbol{h_i}$ represents the $i^{\text{th}}$ timestep hidden activity from the last layer. 
\subsection{Transformer Network}
\label{ssec:subhead}
Attention is a very well known concept in deep learning, specifically for temporal sequences. These models are widely used in speech and NLP applications due to their ability to attend and select relevant features that are useful for solving a particular task. Recently, Multi-head self-attention has become one of the widely used models in NLP and Speech due to its ability to attend different parts of the input using multiple attention heads. Various attention models have been invented over the past, including dot product attention, locality-sensitive attention, additive attention, and so on. In this work, we use a specific type of attention called multi-head self-attention. A detailed multi-head self-attention block is shown in Fig. 2.

\begin{figure}[t]
  \centering
  \includegraphics[width=\linewidth]{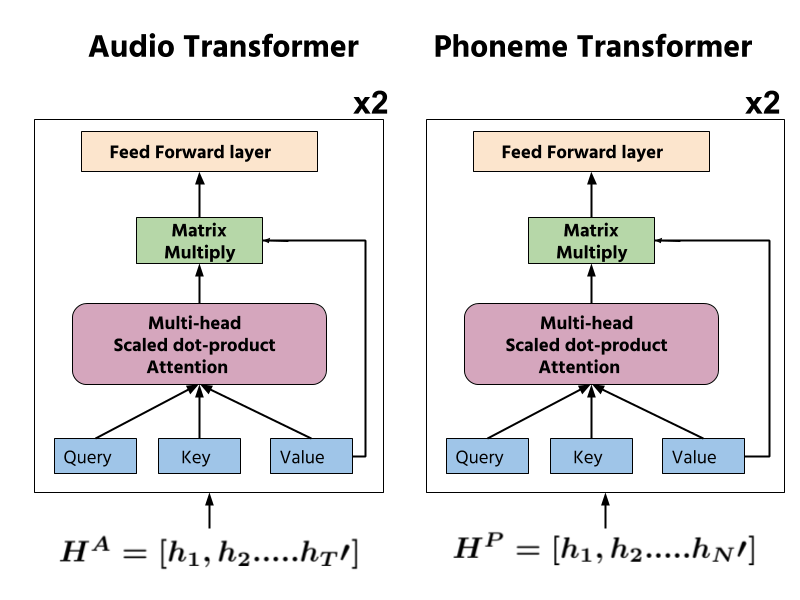}
  \caption{ Multi-head self-attention: Audio Transformer (left) and Phoneme Transformer (right)}
  \label{fig:speech_production}
\end{figure}

The multi-head attention (MHA) layer consists of multihead scaled dot product attention, matrix multiply, and position-wise feed-forward layer. The MHA block consists of M linear layers for query Key and Value matrices, where M is the number of heads in the multi-head attention.
It takes feature representations from Bi-LSTM layer and applies linear transform to create $\boldsymbol{Q_i}$, $\boldsymbol{K_i}$ and $\boldsymbol{V_i}$ using $\boldsymbol{i^{\text{th}}}$ linear transform where, $\boldsymbol{i=[1,2.....M]}$  and M is the total number of attention heads.
The $\boldsymbol{Q_i}$, $\boldsymbol{K_i}$ and $\boldsymbol{V_i}$ are fed into scaled dot product attention layer followed by matrix multiplication between the value matrix and attention weights. The scaled dot product attention $\boldsymbol{A_i}$ for $\boldsymbol{i^{\text{th}}}$ head is defined as follows.

\begin{equation}
\boldsymbol{A_i} = \textsf{Softmax}(\boldsymbol{\frac{Q_iK_i}{d_q}})\boldsymbol{V_i}
\end{equation}

Where $\boldsymbol{d_q}$ is the dimension of the query vector. We combine the attention output from all the heads using simple concatenation and feed them into the feed-forward layer.

\begin{equation}
\boldsymbol{A} = \textsf{Concat}(\boldsymbol{A_1,A_2,A_3...A_i.....A_M}){W_0}
\end{equation}

Our proposed model consists of Audio Transformer and Phoneme Transformer. The Audio Transformer takes output feature representation $\boldsymbol{H^A=[h_1,h_2.....h_T\prime_{}]}$ from Audio Encoder as input, while Phoneme Transformer takes output feature representation $\boldsymbol{H^P=[h_1,h_2.....h_N\prime_{}]}$ from Phoneme Encoder as input. The Query, Key, and Value are obtained using $\boldsymbol{H^A}$ for Audio Transformer. Similarly, the Query, Key, and Value are obtained using $\boldsymbol{H^P}$ for Phoneme Encoder. Both Audio Transformer and Phoneme Transformer consist of 2 layers of Multi-head self-attention with four attention heads each.

\subsection{Statistics Pooling}
\label{ssec:subhead}
The statistics pooling layers pools the frame-level features from both Audio Transformer and Phoneme Transformer. It computes the mean and standard deviation across time in a feature sequence to generate a single utterance-level feature representation from Audio Network and Phoneme Network. The mean and standard deviation vectors are concatenated to obtain the second-order statistics of Audio Encoder and Phoneme Encoder.

\begin{equation}
\boldsymbol{U^A} = \textsf{Concat}({\textsf{mean}(\boldsymbol{A^a})},{\textsf{std}(\boldsymbol{A^a})})
\end{equation}

\begin{equation}
\boldsymbol{U^P} = \textsf{Concat}({\textsf{mean}(\boldsymbol{A^p})},{\textsf{std}(\boldsymbol{A^p})})
\end{equation}

Where, $\boldsymbol{A^a}$, $\boldsymbol{A^p}$ are the outputs from Audio Transformer and Phoneme Transformer respectively. $\boldsymbol{U^A}$ is a Utterance-level feature vector from $\boldsymbol{A^a}$ and $\boldsymbol{U^P}$ is a Utterance-level feature vector from $\boldsymbol{A^p}$ . Both $\boldsymbol{U^A}$ and $\boldsymbol{U^P}$ are having same dimension.

\subsection{Early Fusion}
\label{ssec:subhead}
The utterance-level feature vectors $\boldsymbol{U^A}$ and $\boldsymbol{U^P}$ are the representation learned from different modalities. The early fusion block combines these two representations to combine information from both audio and phoneme modality using simple concatenation. 

\begin{equation}
\boldsymbol{C} = \textsf{Concat}(\boldsymbol{U^A},\boldsymbol{U^P})
\end{equation}

Where $\boldsymbol{C}$ represents the fused feature representation.

\section{Dataset}
\label{ssec:subhead}
\subsection{Speech Recognition dataset}
The proposed model uses both audio features and phoneme sequences to train multi-modal architecture for code-switching identification. We first build LVSCR speech recognition models for Telugu, Tamil, and Gujarati languages to extract the code-switching dataset's phoneme sequences. We use the dataset provided for Low resource speech recognition challenge by Microsoft. The dataset consists of ~40 hours of audio data for each of the languages. All the audio files for each language are sampled at 16KHz to train speech recognition models. The dataset contains both training evaluation split for each language. We first train the GMM-HMM-based speech recognition model for each language using the dataset split. We then evaluate each model with corresponding tests split from their respective language. 

\subsection{Code-switching dataset}
In this section, we discuss the code-switching dataset in detail.  One of the shared tasks' main goal is to detect if an audio file is monolingual or code-switched. The shared task contains two subtasks, 1) Utterance-level identification of monolingual vs. code-switched utterances, and 2)Frame-level identification of language in a code-switched utterance. We focus on subtask-1 in this paper. The dataset for subtask-1 contains 3 Indic languages, Telugu-English, Tamil-English, and Gujarati-English. Each language is provided with training and testing sets. Both training and test sets contain information about whether an audio file is monolingual or code-switched. The statistics of the dataset is given in Table 2. To use our approach, we need a phoneme sequence for every utterance in the code-switching identification dataset. To obtain the phoneme sequence, we use the prebuilt GMM-HMM speech recognition models for their corresponding languages. We use Telugu, Tamil, and Gujarati speech recognition models to extract phoneme sequences for Telugu-English, Tamil-English, and Gujarati-English code-switched dataset, respectively.

 \begin{table}
\centering
  \label{tab:tasks}
  
  \begin{adjustbox}{max width=\textwidth}
  \begin{tabular}{|l|l|l|l|l|l}
    \hline
    \multirow{2}{*}{Dataset} &
     \multicolumn{2}{c}{Train} &
     \multicolumn{2}{c}{Evaluation} \\
     & Duration & Utterances & Duration & Utterances\\
     \hline
     Gujarati & 31.59 & 16780 & 3.59 & 2091 \\
     \hline
     Telugu & 31.59 & 16991 & 4.0 & 2135 \\
     \hline
     Tamil & 30.24 & 10933 & 7.312 & 2642 \\
     \hline
  \end{tabular}
  \end{adjustbox}
  \caption{Code-switching dataset: train and evaluation splits for different languages}
\end{table}

\section{Experiments}
\subsection{Speech Recognition}
Our speech recognition dataset consists of ~40hrs of labeled audio data for 3 Indic languages Telugu, Tamil, and Gujarati. We extract 13-dimensional MFCC features for every 25ms window with a 10ms shift. We use these features to train the acoustic models. We build 3 different speech recognition models, \textit{Telugu-ASR},\textit{Tamil-ASR}, \textit{Gujarati-ASR} for Telugu, Tamil and Gujarati language respectively. The acoustic model for \textit{Telugu-ASR} is trained using $\sim$40 hours of audio data containing $\sim$44K utterances. We train the Bi-Gram language model using training utterances. Similarly, the acoustic models for \textit{Tamil-ASR}, \textit{Gujarati-ASR} are trained using $\sim$40hours of data for each language, and the dataset contains $\sim$39K utterance for Tamil and $\sim$22K. The lexicon for \textit{Telugu-ASR}, \textit{Tamil-ASR} and Gujarati have $\sim$48K, $\sim$58K and $\sim$43K words respectively. We evaluate speech recognition models on $\sim$5 hours of evaluation data for each language.We use Kaldi \footnote{www.kaldi.net} [34] to train speech recognition models. We use KenLM\footnote{www.kenLM.com} [35] to train the language models.

\subsection{Code-switching Identification}
Our proposed approach consists of 2 streams, Audio Network and Phoneme Network. The input to the audio network is a sequence of 13-dimensional MFCC features. We use 25ms window length with 10ms hop length to extract the features.  The Audio Encoder has two convolutional neural layers containing [1x7,1x5] kernels and 64 filters each. Each convolutional layer has a stride of 3. Each of these convolutional layers consists of 1D convolution operation, 1D-batch normalization, and Relu activation function. The audio encoder's convolutional layers help reduce the temporal dimension of the input data and extract higher-level features. The convolutional block's output from the audio encoder is fed to a Bi-LSTM with a hidden size of 128 and a dropout of 0.3.
Similarly, the phoneme encoder has two 1D convolutional layers with kernel size [1x3,1x5] and 64 filters. Each convolution layer has stride 1. The input to the text encoder is a sequence of phoneme embeddings of dimension 128 and is learned as part of the training. The output of the Phoneme Encoder goes to a Bi-LSTM with a hidden layer size of 128 and a dropout of 0.3.  Both Audio transformer and Phoneme Transformer contains two layers of multi-head attention(MHA) blocks. Each MHA block uses four attention heads layer-normalization before scaled dot product operation. The forward feed layer inside MHA has a hidden dimension is 256. We use Adam optimizer [36] with a scheduled learning rate. We use Pytorch [37] framework for implementation. All our models are trained on RTX 2080Ti Graphics cards.

\section{Results}
\label{sec:print}
In this section, we describe the experimental results in detail. We first evaluate our speech recognition system performance for Telugu, Tamil, and Gujarati using standard evaluation data. We then compare our multi-modal approach with uni-modal approaches. We finally examine the effectiveness of transformers for code-switching identification task. We train GMM-HMM-based speech recognition on 40hrs of audio data for each language and evaluate the word error rates on the corresponding test sets. We obtain, $\sim$24\%, $\sim$20\% and $\sim$17\% WER for Telugu, Tamil and Gujarati respectively.

 We train our multi-modal model for Telugu-English, Tamil-English, and Gujarati-English on the code-switching dataset. We evaluate the models on the evaluation dataset for each language. The comparison between multi-modal and uni-modal approaches are shown in Table 2. The experiment \textit{Multi-Modal} represents our approach, where we use both Audio network and Phoneme Network. \textit{Uni-Modal-Audio} represents the model where the classifier is trained using only Audio Network with MFCC features as inputs. Similarly, \textit{Uni-Modal-Phoneme} is the model where the classifier is trained using only Phoneme Network with Phoneme sequences as inputs. It can be shown that our approaches outperform the uni-modal approaches. Our method outperforms the \textit{Uni-Modal-Audio} approach by 6.8\%,7.75\%, and 4.14\% for Telugu, Tamil, and Gujarati, respectively. Similarly, Our method outperforms the \textit{Uni-Modal-Phoneme} approach by 5.46\%,5.47\%, and 13.85\% for Telugu, Tamil, and Gujarati, respectively.

\begin{table}[!htbp]
  \centering
  \label{tab:tasks}
  \begin{adjustbox}{max width=\textwidth}
  
    \begin{tabular}{cc} 
      \hline
      \textbf{System} & \textbf{Accuracy}\\
      \hline
      \textbf{Telugu-English}\\

      \textit{Multi-Modal} & \textbf{85.71}\%\\
      \textit{Uni-Modal-Audio} & 78.91\%\\
      \textit{Uni-Modal-Phoneme} & 80.25\%\\
      \hline
      \textbf{Tamil-English}\\

      \textit{Multi-Modal} & \textbf{86.02}\%\\
      \textit{Uni-Modal-Audio} & 78.27\%\\
      \textit{Uni-Modal-Phoneme} & 80.55\%\\
      \hline
       \textbf{Gujarati-English}\\
      \textit{Multi-Modal} & \textbf{88.85}\%\\
      \textit{Uni-Modal-Audio} & 84.71\%\\
      \textit{Uni-Modal-Phoneme} & 75.80\%\\
      \hline
    \end{tabular}
 \end{adjustbox}
\caption{Comparison of Multi-modal and Uni-Modal approaches. (Bold indicates 
  the best performance)}

\end{table}

We experiment to see the effectiveness of the transformer for code-switching identification task in multi-modal settings. We remove the Audio Transformer and Phoneme Transformer from the original model to see the code-switching identification task's performance. We call this experiment  \textit{CNN+Bi-LSTM}. Similarly, we remove Bi-LSTM from the original architecture to see the effect of using Bi-LSTM in the system performance. The experiment is known as  \textit{CNN+Transformer}. In the final experiment, we keep all the blocks from the original architecture, and the experiment is called \textit{CNN+Bi-LSTM+Transformer}. We conduct these experiments for all the language pairs, and the results are shown in Table 3. It can be seen that the \textit{CNN+Bi-LSTM+Transformer} model is shown to be the best model. It can be seen that the transformers help in improving overall system performance.

\begin{table}[!htbp]
  \centering
  \label{tab:tasks}
  \begin{adjustbox}{max width=\textwidth}
    \begin{tabular}{cc}
      \hline
      \textbf{System} & \textbf{Accuracy}\\
      \hline
      \textbf{Telugu-English}\\
      \textit{CNN+Bi-LSTM} & 85.00\%\\
      \textit{CNN+Transformer} & 85.30\%\\
      \textit{CNN+Bi-LSTM+Transformer} &  \textbf{85.71}\%\\
      \hline
       \textbf{Tamil-English}\\
      \textit{CNN+Bi-LSTM} & 85.70\%\\
      \textit{CNN+Transformer} & \textbf{86.02}\%\\
      \textit{CNN+Bi-LSTM+Transformer} & 85.60\%\\
      \hline
       \textbf{Gujarati-English}\\
      \textit{CNN+Bi-LSTM} & 88.01\%\\
      \textit{CNN+Transformer} & 87.90\%\\
      \textit{CNN+Bi-LSTM+Transformer} & \textbf{88.85}\%\\
      \hline
    \end{tabular}
  \end{adjustbox}
  \caption{Comparison of Different neural architectures. Bold indicates 
  the best performance}
\end{table}

\section{Conclusions}
Code-switching detection is one of the most challenging and still unsolved problems in the speech processing field. This work proposes a new approach for improving code-switching detection performance using multi-modal learning combined with transformer networks by utilizing the phonetic information. Our method consists of two streams neural network where one stream processes audio features like MFCC while the other stream processes phoneme sequences. Multi-modal learning helps capture information shared between different modalities to learn a better code-switching identification model. We also show that transformer models help improve code-switching detection performance due to their ability to attend and select features using a self-attention mechanism for better code-switching detection performance.
Our experiments show that our multi-modal learning approach outperforms the uni-modal approaches on Microsoft's code-switching identification dataset.

\section{Acknowledgements}
We want to thank Microsoft India for providing the dataset for this research work. Any opinion, findings, conclusions, or recommendations expressed in this material are those of the author(s) and do not necessarily reflect the views of Freshworks Inc.

\end{document}